\documentclass[%
reprint,
superscriptaddress,
 amsmath,amssymb,
floatfix,
]{revtex4-1}

\usepackage[dvipdfmx]{graphicx}
\usepackage{dcolumn}
\usepackage{float}
\usepackage[dvipsnames]{xcolor}
\usepackage[breaklinks]{hyperref}

\newcommand{\mri}{\mathrm{i}} 
\newcommand{\mrd}{\mathrm{d}} 

\begin{document}


\title{Ground-state phase diagram of spin-$S$ Kitaev-Heisenberg models}

\author{Kiyu Fukui}
\email{k.fukui@aion.t.u-tokyo.ac.jp}
\affiliation{Department of Applied Physics, The University of Tokyo, Bunkyo, Tokyo 113-8656, Japan}
\author{Yasuyuki Kato}%
\affiliation{Department of Applied Physics, The University of Tokyo, Bunkyo, Tokyo 113-8656, Japan}
\author{Joji Nasu}
\affiliation{Department of Physics, Tohoku University, Sendai, Miyagi 980-8578, Japan}
\affiliation{PRESTO, Japan Science and Technology Agency, Honcho Kawaguchi, Saitama 332-0012, Japan}

\author{Yukitoshi Motome}
\affiliation{Department of Applied Physics, The University of Tokyo, Bunkyo, Tokyo 113-8656, Japan}




\date{\today}

\begin{abstract}
The Kitaev model, whose ground state is a quantum spin liquid (QSL), was originally conceived for spin $S=1/2$ moments on a honeycomb lattice. In recent years, the model has been extended to higher $S$ from both theoretical and experimental interests, but the stability of the QSL ground state has not been systematically clarified for general $S$, especially in the presence of other additional interactions, which inevitably exist in candidate materials. Here we study the spin-$S$ Kitaev-Heisenberg models by using an extension of the pseudofermion functional renormalization group method to general $S$. We show that, similar to the $S=1/2$ case, the phase diagram for higher $S$ contains the QSL phases in the vicinities of the pristine ferromagnetic and antiferromagnetic Kitaev models, in addition to four magnetically ordered phases. We find, however, that the QSL phases shrink rapidly
with increasing $S$, becoming vanishingly narrow for $S\geq 2$, whereas the phase boundaries between the ordered phases remain almost intact. Our results provide a reference for the search of higher-$S$ Kitaev materials.
\end{abstract}

\maketitle

\section{Introduction}\label{sec:intro}
Since the proposal by Anderson~\cite{Anderson1973, Fazekas1974}, the quantum spin liquid (QSL),
which is a quantum disordered state in magnets with fascinating features such as quantum entanglement and fractional exciations, has been studied intensively from both theoretical and experimental points of view~\cite{Diep, Balents2010, Lacroix, Zhou2017}. 
Despite the long history of research, well-established examples of the QSL are limited, and the realization of the QSL in most of the candidate models and materials is still under debate. The Kitaev model has brought a breakthrough, by providing a rare example of exact QSLs in more than one dimension~\cite{Kitaev2006}. This is a frustrated quantum spin model with $S=1/2$ moments on a two-dimensional honeycomb lattice. Despite strong frustration arising from the bond-dependent anisotropic interactions, the model is exactly solvable, and it is proved that the ground state is a QSL with fractional excitations of itinerant Majorana fermions and localized $Z_2$ fluxes. Since the feasibility of the model was proposed for spin-orbit coupled Mott insulators~\cite{Jackeli2009}, a number of intensive searches for the candidate materials have been carried out from both experimental and theoretical perspectives~\cite{rau2016, Trebst2017, Winter2017, Takagi2019, Motome2020a, Motome2020, Trebst2022}. The representative examples of the candidate materials are Na$_2$IrO$_3$~\cite{Chaloupka2010, Singh2010, Singh2012, Comin2012, Chaloupka2013, Foyevtsova2013, Sohn2013, Katukuri2014, Yamaji2014, HwanChun2015, Winter2016}, $\alpha$-Li$_2$IrO$_3$~\cite{Singh2012, Chaloupka2013, Winter2016}, and $\alpha$-RuCl$_3$~\cite{Plumb2014, Kubota2015, Winter2016, Yadav2016, Sinn2016}.  
In recent years, a number of new candidates have been proposed, such as cobalt compounds~\cite{Liu2018, Sano2018, Liu2020, Kim2022}, iridium ilmenites~\cite{Haraguchi2018, Haraguchi2020, Jang2021}, and $f$-electron compounds~\cite{Jang2019, Xing2020, Jang2020, Ramanathan2021, Daum2021}. 

Although the Kitaev model was originally introduced for the $S=1/2$ moments, its higher-spin generalization has also attracted attention. 
For instance, 
as the $S=1$ candidates, the possibility of the Kitaev-type anisotropic interactions in $A_{3}$Ni$_2X$O$_6$ ($A$ = Li and Na, $X$ = Bi and Sb) was discussed theoretically~\cite{Stavropoulos2019}, and later, the inelastic neutron scattering measurement on Na$_2$Ni$_2$TeO$_6$ indicated the pronounced effect of the ferromagnetic (FM) Kitaev interaction~\cite{Samarakoon2021}. In addition, as the $S=3/2$ case, the importance of the Kitaev interaction was claimed from the Hall micromagnetometry measurement for CrBr$_3$~\cite{Kim2019} and the ferromagnetic resonance experiment for CrI$_3$~\cite{Lee2020a}.
The density functional theory calculation also implies the potential Kitaev QSL in CrI$_3$~\cite{Xu2018}. Moreover, combining the density functional theory and the exact diagonalization (ED) calculation, the realization of the antiferromagnetic (AFM) Kitaev QSL was predicted for epitaxially strained monolayers of CrSiTe$_3$ and CrGeTe$_3$ with $S=3/2$ moments~\cite{Xu2018, Xu2020}. Meanwhile, extensions of the Kitaev model to general $S$ have been studied intensively in recent years~\cite{Baskaran2008, Suzuki2018, Koga2018, Oitmaa2018, Minakawa2019, Koga2020, Zhu2020, Hickey2020, Lee2019, Khait2021, Jin2021a, Chen2022, Bradley2022}. 
It was proved that the ground state of the models is a QSL state for arbitrary $S$, where the spin correlations vanish beyond nearest neighbors~\cite{Baskaran2008}.
The stability of QSLs in such higher-$S$ generalization, however, has not been systematically clarified in the presence of other additional interactions, such as the Heisenberg interaction~\cite{Chaloupka2010, Chaloupka2013} and the off-diagonal symmetric interactions~\cite{Rau2014, Katukuri2014, Chaloupka2015}, which inevitably exist in real materials, because of the increase of the Hilbert space with $S$.

In this paper, we present our numerical results on the ground state of higher-spin generalization of the Kitaev model including the Heisenberg interaction, dubbed the spin-$S$ Kitaev-Heisenberg model, by using an extension of the pseudofermion functional renormalization group (PFFRG) method~\cite{Reuther2009, Reuther2010, Baez2017}. 
Performing the calculations for the models with $S=1$, $3/2$, $2$, $5/2$, and $50$ in addition to $S=1/2$,
we elucidate the ground-state phase diagram by systematically changing $S$ and the ratio between the Kitaev and Heisenberg interactions.  
Our results for $S=1/2$ and $1$ are consistent with the previous studies~\cite{Chaloupka2013, Dong2020}: there are QSL phases around the pure Kitaev cases without the Heisenberg interaction in both FM and AFM cases, in addition to four magnetically ordered phases, the FM, N\'{e}el AFM, zigzag AFM, and stripy AFM phases. 
We also show that the result for $S=50$ is also consistent with that for the classical spins corresponding to $S=\infty$~\cite{Price2013}. 
Combining them with the results for $S=3/2$, $2$, and $5/2$, we find that while the phase boundaries between the magnetically ordered phases are almost unchanged, the two QSL regions shrink rapidly with increasing $S$, and appear to be very fragile against the introduction of the Heisenberg interaction for the cases with $S\geq 2$.

The structure of this paper is as follows. In Sec.~\ref{sec:model}, we introduce the spin-$S$ Kitaev-Heisenberg model and briefly review the previous studies for $S=1/2$, $1$, and classical spins. In Sec.~\ref{sec:method}, we describe the PFFRG method and its extension to general $S$, and present the conditions of our numerical calculations. We present our results for the ground-state phase diagram in Sec.~\ref{subsec:phase_diagram}. Then, we show the results for the pure Kitaev cases and in their vicinities in Secs.~\ref{subsec:Kitaev_cases} and \ref{subsec:vicinity}, respectively, and those for the four magnetically ordered states in Sec.~\ref{subsec:ordered}. Finally, we summarize our main findings in Sec.~\ref{sec:summary}.

\section{Model}\label{sec:model}

\begin{figure}
    \centering
     \includegraphics[width=0.9\columnwidth,clip]{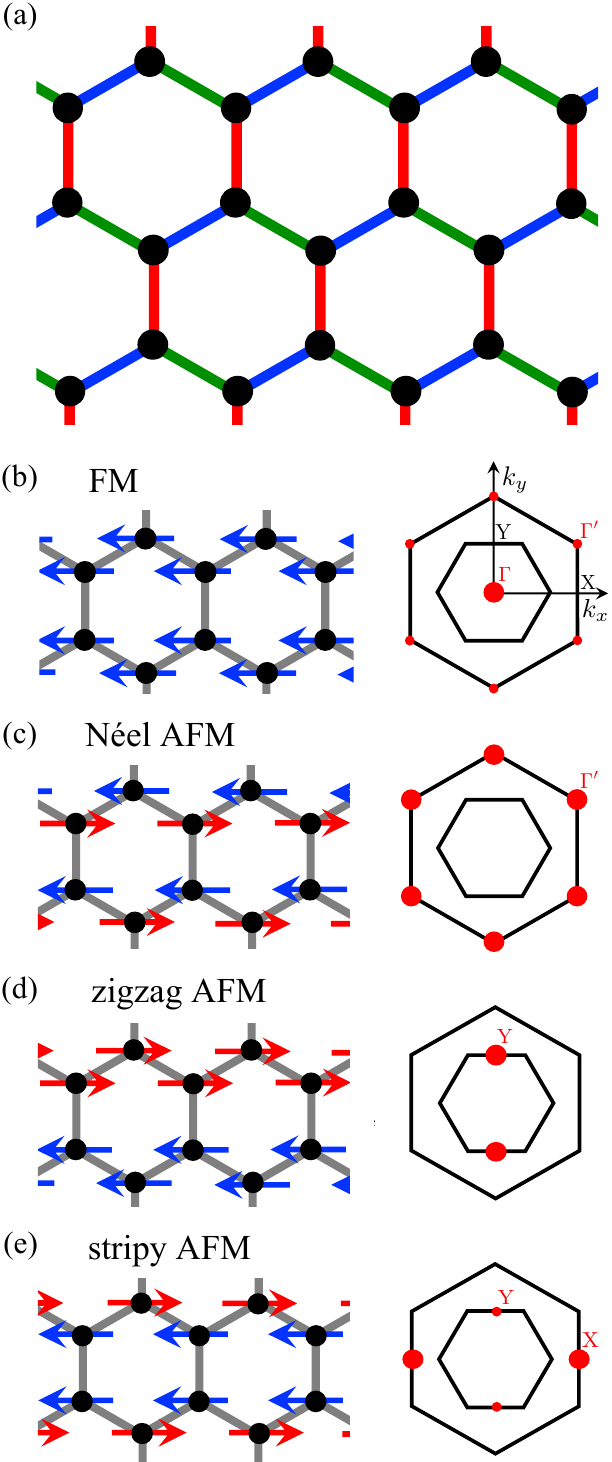}
    \caption{(a) Schematic picture of the spin-$S$ Kitaev-Heisenberg model defined on the honeycomb lattice. The blue, green, and red bonds represent the $\mu=x$, $y$, and $z$ bonds in Eq.~\eqref{eq:KH_model}, respectively. (b)--(e) Spin configurations (left) and the corresponding peak positions of the spin susceptibility in momentum space (right) for four magnetically ordered states in the spin-$S$ Kitaev-Heisenberg model. In the right panels, the peak positions are indicated by the filled red circles, whose sizes represent the peak intensities schematically. The inner hexagon indicates the first Brillouin zone, while the outer one indicates the Brillouin zone up to the third one. $\Gamma$, $\mathrm{X}$, $\mathrm{Y}$, and $\Gamma'$ indicate the high symmetry points.
}
    \label{fig:KH_model}
\end{figure}

We study higher-spin generalization of the Kitaev-Heisenberg model on the honeycomb lattice as a minimal model for higher-spin candidate materials. The Hamiltonian is given by
\begin{align}\label{eq:KH_model}
    \mathcal{H}&=\frac{1}{(2S)^{2}}\sum_{\mu=x,y,z}\sum_{\langle i,j\rangle_\mu}\left[ 2KS^{\mu}_iS^{\mu}_j + J\mathbf{S}_i\cdot\mathbf{S}_j\right],
\end{align}
where the summation of $\langle i, j\rangle_{\mu}$ runs over pairs of nearest-neighbor sites $i$ and $j$ connected by $\mu$ bond, and $S^{\mu}_i$ is the $\mu$ component of the spin-$S$ operator at site $i$: $\mathbf{S}_i=(S^x_i,\ S^y_i,\ S^z_i)$. A schematic picture of the model is shown in Fig.~\ref{fig:KH_model}(a), in which the  
$x$, $y$, and $z$ bonds are represented by blue, green, and red, respectively. The first and second terms in Eq.~\eqref{eq:KH_model}  represent the Kitaev and Heisenberg interactions, respectively;
$K$ and $J$ are the coupling constants parametrized as
\begin{equation}\label{eq:KandJ}
    K=\sin(2\pi\xi),
\quad J=\cos(2\pi\xi),
\end{equation}
by using the parameter $\xi\in[0, 1]$. We take the energy unit as $\sqrt{K^2+J^2}=1$. 
 
According to Eq.~\eqref{eq:KandJ}, the Kitaev coupling is FM ($K<0$) for $\frac{1}{2}<\xi<1$, while it is AFM ($K>0$) for $0<\xi<\frac{1}{2}$. Meanwhile, the Heisenberg interaction is FM ($J<0$) for $\frac{1}{4}<\xi<\frac{3}{4}$, while AFM ($J>0$) for $0\leq\xi<\frac{1}{4}$ and $\frac{3}{4}<\xi\leq1$. There are four special values of $\xi$: $\xi=0$, $\frac14$, $\frac12$, and $\frac34$. When $\xi=\frac{3}{4}$ and $\frac{1}{4}$, the Heisenberg interaction vanishes ($J=0$), and the Hamiltonian in Eq.~\eqref{eq:KH_model} becomes the FM and AFM Kitaev models, respectively, where the ground states are QSLs for arbitrary $S$ as described below.
Meanwhile, when $\xi=\frac{1}{2}$ and $0$, the Kitaev interaction vanishes ($K=0$), and the Hamiltonian corresponds to the FM and AFM Heisenberg models, respectively, for which the system has the SU($2$) symmetry, and the FM and N\'{e}el AFM orders are realized in the ground state. In addition, owing to the four-sublattice transformation~\cite{Khaliullin2005, Chaloupka2010, Chaloupka2013, Chaloupka2015}, there are two more hidden SU($2$) points at $\xi=\frac{7}{8}$ and $\frac{3}{8}$ corresponding to $\xi=\frac12$ and $0$, respectively. 

This model with $S=1/2$ was introduced as an effective model for the candidate materials like Na$_2$IrO$_3$ and $\alpha$-Li$_2$IrO$_3$, and the phase diagram was calculated by various methods, such as the ED method~\cite{Chaloupka2010,Chaloupka2013}, the density matrix renormalization group (DMRG)~\cite{Jiang2011}, the slave-particle mean-field approximation~\cite{Schaffer2012}, and the tensor network method~\cite{Iregui2014}, the cluster mean-field approximation~\cite{Gotfryd2017}, the high-temperature expansion~\cite{Singh2017}, and the quantum Monte Carlo method~\cite{Sato2021}. 
The PFFRG calculation was also performed for $\frac{3}{4}\leq\xi\leq1$~\cite{Reuther2011a}. Through these previous studies, it was shown that the ground-state phase diagram of the spin-$1/2$ Kitaev-Heisenberg model contains four magnetically ordered phases, FM, N\'eel AFM, zigzag AFM, and stripy AFM phases, in addition to the QSL phases extended from the two pure Kitaev cases as $\xi=\frac34$ and $\frac14$ where the ground states are the exact QSLs. The schematic spin configurations of the ordered states are shown in Figs.~\ref{fig:KH_model}(b)--\ref{fig:KH_model}(e).

The extensions of the model to higher $S$ have also been studied. For the pure Kitaev cases at $\xi=\frac34$ and $\frac14$, it was analytically shown that the spin correlations vanish beyond nearest neighbors for arbitrary $S$~\cite{Baskaran2008}. 
This means the ground states are QSLs for arbitrary $S$.
In the presence of the Heisenberg interaction, the ground-state phase diagram for the $S=1$ case was studied by the ED method~\cite{Stavropoulos2019} and DMRG~\cite{Dong2020}. 
The results contain the four magnetically ordered phases and the QSL phases similar to the $S=1/2$ case, but the QSL regions are narrower than those for $S=1/2$, reflecting the suppression of quantum fluctuations by increasing $S$. 
The model was also studied for the classical spins, which corresponds to $S=\infty$, by classical Monte Carlo (MC) simulation~\cite{Price2012, Price2013}. 
The phase diagram also shows four magnetically ordered phases, except for 
the pure Kitaev cases at $\xi=\frac34$ and $\frac14$, where the classical ground state is disordered with macroscopic degeneracy and the spin excitation has zero modes~\cite{Baskaran2008, Chandra2010, Price2013}. 
The results by systematically changing $S$ are unknown to the best of our knowledge, which we investigate in the following by the PFFRG method.

\section{method}\label{sec:method}
To elucidate the ground-state phase diagram of the spin-$S$ Kitaev-Heisenberg model in Eq.~\eqref{eq:KH_model}, we use an extension of the PFFRG method to general $S$~\cite{Baez2017,Buessen2018,Iqbal2019}. The PFFRG provides a powerful numerical method for frustrated quantum spin systems~\cite{Reuther2009, Reuther2010}, and it has been  applied to a variety of $S=1/2$ systems with the Heisenberg~\cite{Reuther2009, Reuther2010}, $XXZ$~\cite{Gottel2012, Buessen2018}, Kitaev-like~\cite{Reuther2011a, Reuther2012, Reuther2014a, Revelli2019a}, off-diagonal~\cite{Hering2017, Buessen2019}, long-range dipolar~\cite{Keles2018, Keles2018a, Fukui2022}, and SU($2$)$\times$SU($2$) interactions~\cite{Kiese2019}.  
In this method, the $S=1/2$ spin operator is expressed in terms of auxiliary fermions~\cite{Abrikosov1965}, called pseudofermions, as
\begin{align}\label{eq:pseudofermion}
    S^{\mu}_i=\frac{1}{2}\sum_{\alpha, \alpha'}f^\dagger_{i\alpha'}\sigma^{\mu}_{\alpha',\alpha}f_{i\alpha},
\end{align}
where $f_{i\alpha}$ ($f^\dagger_{i\alpha})$ is an annihilation (creation) operator of the pseudofermion at site $i$ with spin $\alpha\in\{\uparrow,\ \downarrow\}$, and $\sigma^{\mu}$ is the $\mu$ ($=x$, $y$, or $z$) component of the Pauli matrices. Here and hereafter, we set the reduced Planck constant $\hbar$ as unity.

The extension to arbitrary spin length $S$ was introduced by preparing $2S$ copies of the spin-$1/2$ moment to represent a spin-$S$ moment as
\begin{equation}\label{eq:spin_S}
    S^{\mu}_i=\sum_{\kappa=1}^{2S}S^{\mu}_{i\kappa}=\frac{1}{2}\sum_{\kappa=1}^{2S}\sum_{\alpha, \alpha'}f^\dagger_{i\alpha'\kappa}\sigma^{\mu}_{\alpha',\alpha}f_{i\alpha\kappa},
\end{equation}
where 
$S^{\mu}_{i\kappa}$ represents the $\mu$ component of the spin-$1/2$ moment at site $i$ in the $\kappa$th copy~\cite{Baez2017}. 
Since this expression enlarges the Hilbert space, one needs a projection to the physical subspace in which the spin moment is maximized at each site. However, it was demonstrated that, in the zero-temperature PFFRG calculations, the spin systems tend to maximize the local moments in the ground state without any projection, even in the presence of frustration~\cite{Baez2017}. 
Following the previous studies~\cite{Baez2017, Buessen2018, Iqbal2019}, we introduce no projection in the following calculations.

By substituting Eq.~\eqref{eq:spin_S} into Eq.~\eqref{eq:KH_model}, we obtain the fermionic Hamiltonian in the quartic form of the pseudofermion operators. In the following, we adopt the fermionic one-particle irreducible FRG~\cite{Salmhofer, Salmhofer2001, Kopietz, Metzner2012, Polonyi, Platt2013} for the resultant quartic Hamiltonian.
In this framework, we solve the fermionic FRG flow equations for the self-energy $\Sigma$ and the two-particle vertex function $\Gamma$ at the level of the one-loop truncation in a fully self-consistent form, which is called the Katanin scheme~\cite{Katanin2004}. 
By exploiting the locality of the pseudofermions due to the absence of the kinetic energy terms and the fact that the coupling constants in the pseudofermion Hamiltonian do not depend on $\kappa$, the flow equations are written as
\begin{align}\label{eq:self_energy_flow}
   &2\pi\delta(\omega_{1'}-\omega_1)\frac{\mathrm{d}}{\mathrm{d}\Lambda}\Sigma^{\Lambda}(\omega_1)\notag\\
=&\sum_{x_2}\mathcal{S}^{\Lambda}(\omega_2)\biggl[\sum_j\Gamma^{\Lambda}_{i_1j}(x_2, x_{1'}; x_1, x_2)\notag\\
   &\qquad \qquad \qquad -\frac{1}{2S}\Gamma^{\Lambda}_{i_1i_1}(x_{1'}, x_2; x_1, x_2) \biggr],
\end{align}
\begin{align}\label{eq:vertex_flow}
    &\frac{\mathrm{d}}{\mathrm{d}\Lambda}\Gamma^{\Lambda}_{i_1i_2}(x_{1'}, x_{2'}; x_1, x_2)\notag\\
=&-\sum_{x_3, x_4}L^{\Lambda}(\omega_3, \omega_4)\notag\\
    &\times\biggl[\frac{1}{2S}\Gamma^{\Lambda}_{i_1i_2}(x_{1'}, x_{2'}; x_3, x_4)\Gamma^{\Lambda}_{i_1i_2}(x_3, x_4; x_1, x_2)\notag\\
    &\quad -\sum_j \Gamma^{\Lambda}_{i_1j}(x_{1'}, x_4; x_1, x_3)\Gamma^{\Lambda}_{ji_2}(x_3, x_{2'}; x_4, x_2)\notag\\
    &\quad +\frac{1}{2S}\Gamma^{\Lambda}_{i_1i_2}(x_{1'}, x_4; x_1, x_3)\Gamma^{\Lambda}_{i_2i_2}(x_{2'}, x_3; x_4, x_2)\notag\\
    &\quad +\frac{1}{2S}\Gamma^{\Lambda}_{i_1i_1}(x_4, x_{1'}; x_1, x_3)\Gamma^{\Lambda}_{i_1i_2}(x_3, x_{2'}; x_4, x_2)\notag\\
    &\quad +\frac{1}{2S}\Gamma^{\Lambda}_{i_1i_2}(x_4, x_{2'}; x_1, x_3)\Gamma^{\Lambda}_{i_1i_2}(x_{1'}, x_3; x_4, x_2)\biggr],
\end{align}
where $\delta(\omega)$ is the delta function, $\Lambda$ denotes the energy cutoff scale in the renormalization group method, and $\Gamma^{\Lambda}_{i_1i_2}$ represents the two-particle vertex function, rescaled as $\Gamma^{\Lambda}_{i_1i_2}=2S\Tilde{\Gamma}^{\Lambda}_{i_1i_2}$~\cite{Baez, Baez2017}, where $\Tilde{\Gamma}^{\Lambda}_{i_1i_2}$ is the unrescaled two-particle vertex function defined in the standard manner~\cite{Salmhofer, Salmhofer2001, Kopietz, Metzner2012, Polonyi, Platt2013},
between pseudofermions on the site $i_1$ and $i_2$; 
$x=(\omega,\ \alpha)$ denotes a set of the Matsubara frequency $\omega$ and the spin index $\alpha$, 
for which the summation is taken as $ \sum_{x}=\int^{\infty}_{-\infty}\frac{\mathrm{d}\omega}{2\pi}\sum_{\alpha}$.
In Eq.~\eqref{eq:self_energy_flow}, $\mathcal{S}^{\Lambda}(\omega)$ represents the single-scale propagator given by 
\begin{equation}
\mathcal{S}^{\Lambda}(\omega)=-\frac{\delta(\lvert\omega\rvert-\Lambda)}{\mri\omega-\Sigma^{\Lambda}(\omega)},
\end{equation}
and $L^{\Lambda}(\omega, \omega')$ in Eq.~\eqref{eq:vertex_flow} is defined as
\begin{equation}
    L^{\Lambda}(\omega, \omega')=\frac{\mrd}{\mrd\Lambda}
\left[G^{\Lambda}(\omega)G^{\Lambda}(\omega')\right],
\end{equation}
where $G^{\Lambda}(\omega)$ is the full propagator given by
\begin{equation}
    G^{\Lambda}(\omega)=\frac{\Theta(\lvert \omega\rvert-\Lambda)}{\mri\omega-\Sigma^{\Lambda}(\omega)},
\end{equation}
with the Heaviside function $\Theta(x)$, working as the cutoff function to project out all modes for $\lvert\omega\rvert<\Lambda$. 
The second term without the factor of $\frac{1}{2S}$ in Eq.~\eqref{eq:vertex_flow} represents contributions from the random phase approximation and plays an important role in the formation of long-range orders. This term becomes dominant in the $S\to\infty$ limit; Eq.~\eqref{eq:vertex_flow} corresponds to the Luttinger-Tisza method for classical spins~\cite{Luttinger1946, Luttinger1951} in this limit~\cite{Baez2017, Baez}.
Note that all the functions in Eqs.~\eqref{eq:self_energy_flow} and \eqref{eq:vertex_flow} are obtained by the summations over $\kappa$, and hence, independent of $\kappa$.
In addition, since the Kitaev-Heisenberg model has only $S^{\mu}_{i}S^{\mu}_{j}$ type spin interactions, the two-particle vertex function can be parametrized as 
\begin{align}
&\Gamma^{\Lambda}_{i_1i_2}(x_{1'}, x_{2'}; x_{1}, x_{2})\notag\\
=&2\pi\delta(\omega_1'+\omega_2'-\omega_1-\omega_2)\notag\\
    &\times
\biggl[\sum_{\mu=x,y,z}\Gamma^{\mu,\Lambda}_{i_1i_2}(s,t,u)\sigma^{\mu}_{\alpha_1',\alpha_1}\sigma^{\mu}_{\alpha_2',\alpha_2}\notag\\
    & 
\qquad \qquad +\Gamma^{\mathrm{d},\Lambda}_{i_1i_2}(s,t,u)\delta_{\alpha_1',\alpha_1}\delta_{\alpha_2',\alpha_2}
\biggr]\label{eq:Gamma_basis},
\end{align}
with
\begin{align}
    s=\omega_1'+\omega_2',
\
    t=\omega_1'-\omega_1,
\
    u=\omega_1'-\omega_2,
\end{align}
where $\Gamma^{\mu,\Lambda}_{i_1i_2}(s, t, u)$ and $\Gamma^{\mathrm{d},\Lambda}_{i_1i_2}(s, t, u)$ represent the renormalized dynamical coupling between the $\mu$ component of the pseudofermion spins and that between the pseudofermion densities, respectively. To solve the integro-differential equations in Eqs.~\eqref{eq:self_energy_flow} and~\eqref{eq:vertex_flow}, we start from the initial conditions given by
\begin{align}
    \Sigma^{\Lambda\to\infty}(\omega)&=0,\\
    \Gamma^{\mu,\Lambda\to\infty}_{i_1i_2}(s, t, u)&=
    \begin{cases}
    (2K+J)/4 & \text{on $\langle i_1,i_2\rangle_{\mu}$}\\
    J/4 & \text{on $\langle i_1,i_2\rangle_{\nu\neq\mu}$}\\
    0 & \text{otherwise}
    \end{cases},\\
    \Gamma^{\mathrm{d},\Lambda\to\infty}_{i_1i_2}(s, t, u)&=0.
\end{align}

To detect magnetic instabilities, we calculate the $z$ components of the spin susceptibility in momentum space defined as
\begin{equation}
\chi^{zz, \Lambda}(\mathbf{k})=\frac{1}{N}\sum_{i, j}\chi^{zz, \Lambda}_{ij}\mathrm{e}^{-\mathrm{i}\mathbf{k}\cdot(\mathbf{r}_{i}-\mathbf{r}_{j})},
\end{equation}
where $N$ is the number of spins, $\mathbf{r}_{i}$ and $\mathbf{r}_{j}$ represent the real-space coordinates of sites $i$ and $j$, respectively; $\chi^{zz, \Lambda}_{ij}$ is the spin susceptibility in real space calculated by the solutions of the flow equations as
\begin{widetext}
\begin{align}\label{eq:suscep}
 \chi^{zz, \Lambda}_{ij}&=\int^{\infty}_0\frac{\mathrm{d}\tau}{2S}\ \langle T_{\tau}S^{z}_i(\tau)S^{z}_j(0)\rangle_{\Lambda}\notag\\ 
    &=-\int^{\infty}_{-\infty}\frac{\mathrm{d}\omega}{4\pi}\ G^{\Lambda}(\omega)^2\delta_{i,j}-\int^{\infty}_{-\infty}\frac{\mathrm{d}\omega\mathrm{d}\omega'}{8\pi^2}\ G^{\Lambda}(\omega)^2G^{\Lambda}(\omega')^2\big[2\Gamma^{z, \Lambda}_{ij}(\omega+\omega', 0, \omega-\omega')-\frac{1}{2S}\big\{\Gamma^{z,\Lambda}_{ii}(\omega+\omega', \omega-\omega', 0)\notag\\
    &\qquad\qquad\qquad\qquad\qquad\qquad\qquad\qquad\qquad\qquad\qquad -\sum_{\mu = x, y}\Gamma^{\mu, \Lambda}_{ii}(\omega+\omega', \omega-\omega', 0)+\Gamma^{\mathrm{d}, \Lambda}_{ii}(\omega+\omega', \omega-\omega', 0)\big\}\delta_{i,j}\big],
\end{align}
\end{widetext}
where $S^{\mu}_{i}(\tau)=\mathrm{e}^{\tau\mathcal{H}}S^{\mu}_{i}\mathrm{e}^{-\tau\mathcal{H}}$, 
$\langle T_{\tau}\cdots\rangle_{\Lambda}$ means the expectation value of the imaginary-time-ordered operators, and $\delta_{i,j}$ is the Kronecker delta. 
Note that the $x$ and $y$ components of the susceptibility are obtained by using the $C_3$ rotational symmetry of the model. 
A magnetic instability is signaled by divergence of $\chi^{zz, \Lambda}(\mathbf{k})$ at a momentum corresponding to the ordering vector, and the critical value of $\Lambda$ is called the critical cutoff scale $\Lambda_{\mathrm{c}}$.
In practice, however, due to the finite system size and the finite frequency grid, the $\Lambda$ dependence of $\chi^{zz, \Lambda}(\mathbf{k})$ shows a kink or cusp rather than the divergence. 
Hence, we use such an anomaly to detect the magnetic instability and estimate $\Lambda_{\mathrm{c}}$. 
Meanwhile, the absence of such an anomaly down to $\Lambda\rightarrow 0$ indicates that the system does not undergo any magnetic instability. 
This suggests the realization of a quantum spin liquid state in the ground state.

In the following numerical calculations, we use the logarithmic frequency grid with 64 positive frequency points between 10$^{-4}$ and 250. We also generate the logarithmic $\Lambda$ grid starting from $\Lambda_{\mathrm{max}}=500$ to $\Lambda_{\mathrm{min}}\simeq 0.02$ by multiplying a factor of $0.96$. In the calculations, we neglect two-particle vertex functions between two sites beyond $10$th neighbors, which corresponds to a finite-size cluster containing $N=166$ lattice sites. 

\section{Result}\label{sec:result}
\subsection{Ground-state phase diagram}\label{subsec:phase_diagram}

\begin{figure}
    \centering
    \includegraphics[width=1.0\columnwidth,clip]{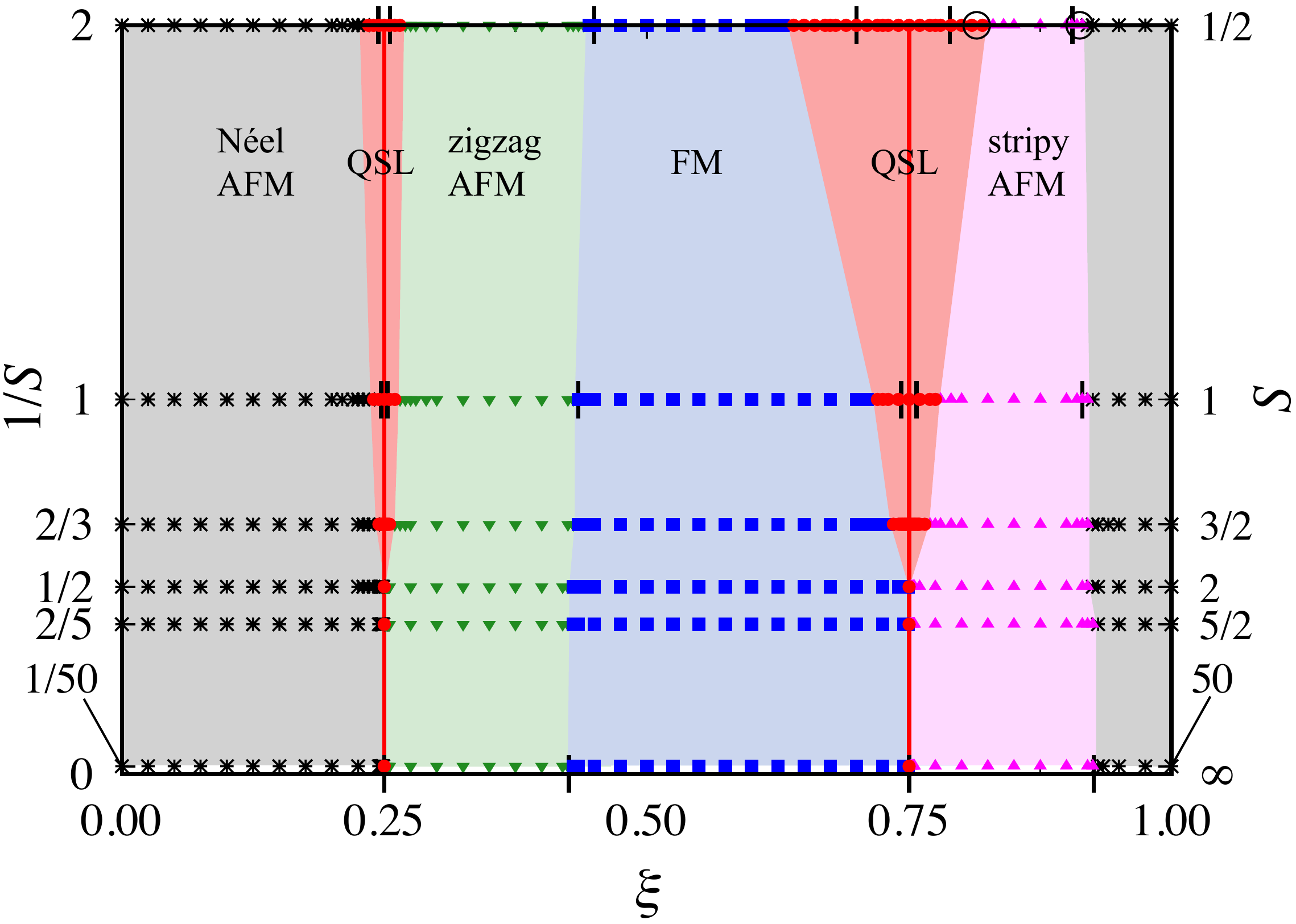}
    \caption{Ground-state phase diagram of the spin-$S$ Kitaev-Heisenberg model in Eq.~\eqref{eq:KH_model} on the plane of $\xi$ and $1/S$. 
The two open circles in the results for $S=1/2$ indicate the phase boundaries obtained by the previous PFFRG study~\cite{Reuther2011a}, while the vertical ticks for $S=1/2$, $1$, and $\infty$ indicate those obtained in the previous studies by the ED~\cite{Chaloupka2013}, DMRG~\cite{Dong2020}, and classical MC calculations~\cite{Price2013}, respectively. 
}
    \label{fig:S_xi_diagram}
\end{figure}

Figure~\ref{fig:S_xi_diagram} summarizes the ground-state phase diagram for the spin-$S$ Kitaev-Heisenberg model obtained by the PFFRG calculations for $S=1/2$ to $5/2$ and $50$. 
First of all, we find QSL-like behavior for the pure Kitaev cases at $\xi=0.25$ and $0.75$ for all the values of $S$. The results of the spin susceptibility are detailed in Sec.~\ref{subsec:Kitaev_cases}. 
Then, around these two cases, we obtain the QSL phases, as indicated by red in Fig.~\ref{fig:S_xi_diagram}. 
The region around the AFM Kitaev case at $\xi=0.25$ is much narrower than that around the FM Kitaev case at $\xi=0.75$, as already known for the cases with $S=1/2$ and $1$. See Sec.~\ref{subsec:vicinity} for the details. 
The phase boundaries obtained by the ED calculation for $S=1/2$~\cite{Chaloupka2013} and the DMRG for $S=1$~\cite{Dong2020} are shown by the vertical ticks in Fig.~\ref{fig:S_xi_diagram}. 
Comparing with these previous studies, our results overestimate the QSL regions. This is presumably due to differences in the numerical methods and the system sizes. 
We note that the phase boundary between the QSL and stripy AFM phases as well as that between the stripy AFM and N\'eel AFM are consistent with the previous PFFRG result for $S=1/2$~\cite{Reuther2011a}, as indicated by the open circle in Fig.~\ref{fig:S_xi_diagram}. 
While further increasing $S$, we find that both QSL regions shrink quickly, and the widths become vanishingly narrow for $S\geq 2$.

In addition to the two QSL phases, we find four magnetically ordered phases for all $S$: N\'eel AFM, zigzag AFM, FM, and stripy AFM, as shown in Fig.~\ref{fig:S_xi_diagram}. 
The schematic pictures of the spin configurations are shown in Fig.~\ref{fig:KH_model}, and the typical behaviors of the spin susceptibility in each region will be presented in Sec.~\ref{subsec:ordered}. 
Our results for the phase boundaries between the ordered phases show good agreement with the previous ones for $S=1/2$ and $1$ indicated by the vertical ticks and the open circle in Fig.~\ref{fig:S_xi_diagram}~\cite{Chaloupka2013,Dong2020,Reuther2011a}. 
They are almost independent of the value of $S$, and the results for $S=50$ coincide well with the previous MC results for $S=\infty$~\cite{Price2013}.

\begin{figure*}
    \centering
    \includegraphics[width=1.0\linewidth,clip]{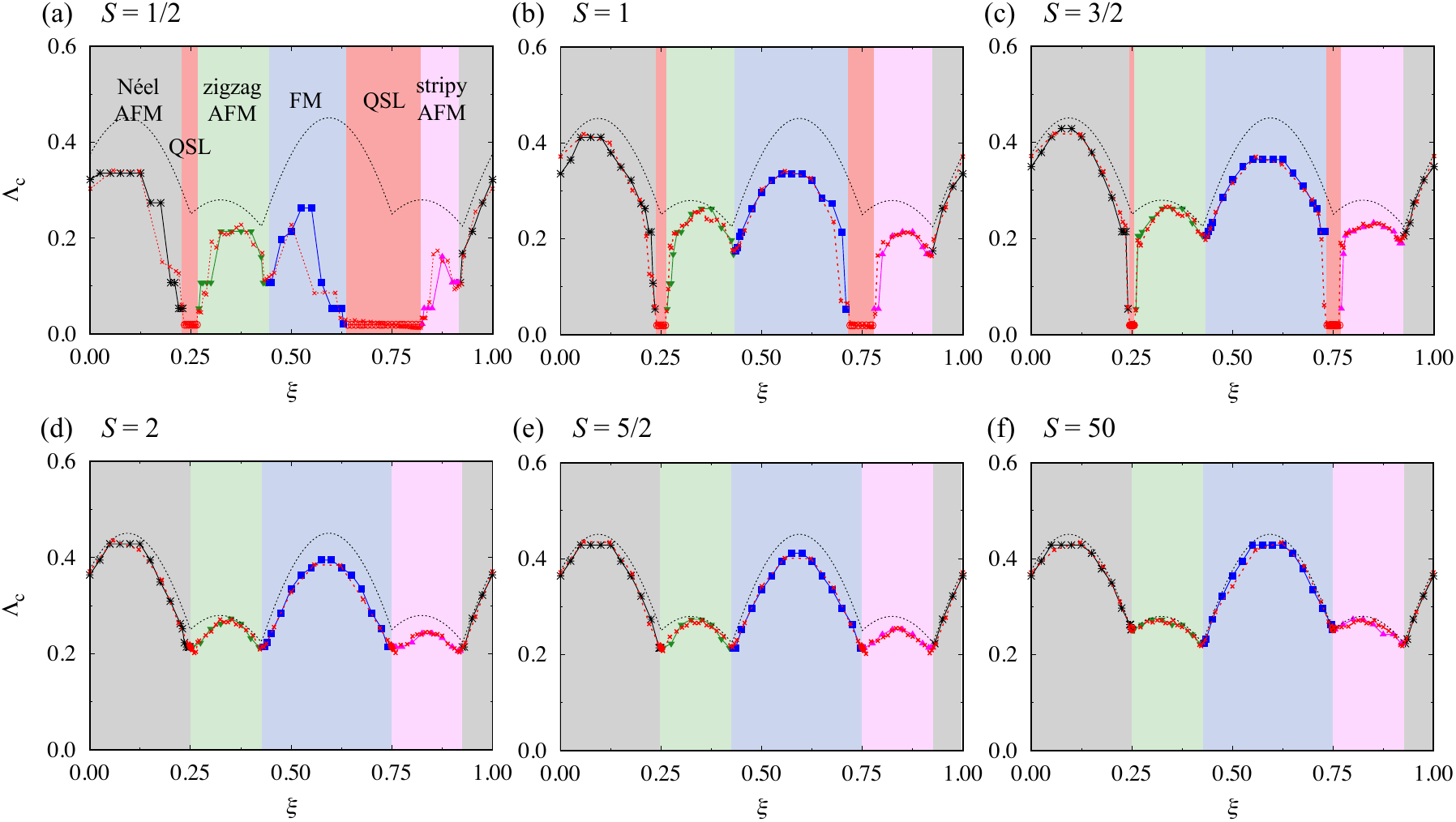}
    \caption{Critical cutoff scale $\Lambda_{\mathrm{c}}$ as a function of $\xi$ for the Kitaev-Heisenberg model with (a) $S=1/2$, (b) $S=1$, (c) $S=3/2$, (d) $S=2$, (e) $S=5/2$, and (f) $S=50$.
The open circles in the QSL regions indicate $\Lambda_{\mathrm{min}}$ down to which no anomalies are found in the spin susceptibility. The filled red circles at $\xi=0.25$ and $0.75$ for $S\geq2$ indicate the characteristic cutoff scale $\Lambda^{*}$ at which the spin susceptibility shows an anomaly; see Sec.~\ref{subsec:Kitaev_cases}.
The black dashed lines denote the absolute value of the energy in the classical limit of $S\to\infty$ obtained by the Luttinger-Tisza method. The red crosses indicate $\Lambda_{\mathrm{c}}$ and $\Lambda^{*}$ obtained by applying the four-sublattice transformation to our results.
}
    \label{fig:Lambda_xi_diagram}
\end{figure*}

The values of the critical cutoff scale $\Lambda_{\mathrm{c}}$, which are determined by anomalies in the spin susceptibility, are summarized in Fig.~\ref{fig:Lambda_xi_diagram}.
In the QSL regions around $\xi=0.25$ and $0.75$, we do not find any anomalies down to $\Lambda_{\mathrm{min}}$ for $S=1/2$, $1$, and $3/2$, as shown in Figs.~\ref{fig:Lambda_xi_diagram}(a)--\ref{fig:Lambda_xi_diagram}(c). 
For $S\geq 2$, however, our results indicate anomalies at nonzero $\Lambda=\Lambda^{*}$ even in the pure Kitaev cases at $\xi=0.25$ and $0.75$, which is presumably an artifact of the current PFFRG calculations; see Sec.~\ref{subsec:Kitaev_cases}.
In the other regions, we obtain nonzero $\Lambda_{\mathrm{c}}$ for all $S$, which shows dome-like $\xi$ dependences in each ordered phase. 
While the $\xi$ dependences are smooth for large $S$, it becomes discrete and stepwise, especially for $S=1/2$. 
This is attributed to the finite frequency grid in our PFFRG calculations, as discussed in Appendix~\ref{appx:stepwise}. 

In Fig.~\ref{fig:Lambda_xi_diagram}, we plot the absolute value of the energy in the classical limit of $S\to\infty$ obtained by the Luttinger-Tisza method for comparison. We find that our estimate of $\Lambda_{\mathrm{c}}$ approaches the classical energy as $S$ increases, which supports that the PFFRG method corresponds to the Luttinger-Tisza method in the $S\to\infty$ limit as mentioned in Sec.~\ref{sec:method}. We also plot the values of $\Lambda_{\mathrm{c}}$ and $\Lambda^{*}$ obtained by applying the four-sublattice transformation~\cite{Khaliullin2005, Chaloupka2010, Chaloupka2013, Chaloupka2015} to our results as follows. In the transformation, $K$, $J$, and $\xi$ in Eq.~\eqref{eq:KandJ} are transformed as
\begin{align}
K'&=\frac{\sin(2\pi\xi)+\cos(2\pi\xi)}{\sqrt{[\sin(2\pi\xi)+\cos(2\pi\xi)]^2+\cos^{2}(2\pi\xi)}},\\
J'&=\frac{-\cos(2\pi\xi)}{\sqrt{[\sin(2\pi\xi)+\cos(2\pi\xi)]^2+\cos^{2}(2\pi\xi)}},\\
\xi'&=\frac{1}{2\pi}\mathrm{arctan}[-\tan(2\pi\xi)-1],
\end{align}
where $K'$ and $J'$ are renormalized to satisfy $K'^{2}+J'^{2}=1$.
In Fig.~\ref{fig:Lambda_xi_diagram}, the red crosses indicate $\Lambda_{\mathrm{c}}$ and $\Lambda^{*}$ at $\xi'$ which are obtained from those at $\xi$ with the same renormalization for $K'$ and $J'$.
Except for the discrete and stepwise behavior for small $S$, we confirm that our results satisfy the four-sublattice symmetry.

Although the current PFFRG calculations are performed at zero temperature, assuming a relation between the energy scale $\Lambda$ and temperature $T$ as $T\simeq\frac{\pi}{2}\Lambda$, which holds for large $T$ and $\Lambda$~\cite{Iqbal2016, Buessen2016, Buessen}, one can regard $\Lambda_{\mathrm{c}}$ as an estimate of the transition temperature $T_{\mathrm{c}}$. 
Indeed, $\Lambda_{\mathrm{c}}$ for $S=50$ qualitatively agrees with the onset temperature of the quasi-long-range order obtained by MC simulation for the classical case~\cite{Price2013}. 
With this assumption, we find that $T_{\mathrm{c}}$ is gradually reduced by quantum fluctuations as $S$ decreases; it is strongly suppressed near the phase boundaries to the QSL, especially for $S\leq 3/2$ where $\Lambda_{\mathrm{c}}\to 0$.
We also find that $T_{\mathrm{c}}$ of the N\'eel AFM and FM phases are larger than those of the zigzag and stripy phases for all $S$. 
We note, however, that $T_{\mathrm{c}}$ is nonzero at the SU($2$) points with $\xi=0.0$ and $0.5$ as well as the hidden SU($2$) points with $\xi=0.375$ and $0.875$ for all $S$, where $T_{\mathrm{c}}$ should be strictly zero because of the Mermin-Wagner theorem~\cite{Mermin1966}. 
This is also an artifact of our PFFRG calculations; indeed, two- or multi-loop extensions beyond the Katanin scheme lead to a better fulfillment of the theorem~\cite{Ruck2018, Thoenniss2020, Kiese2022}.

\subsection{Pure Kitaev cases}\label{subsec:Kitaev_cases}

\begin{figure}
    \centering
    \includegraphics[width=1.0\columnwidth,clip]{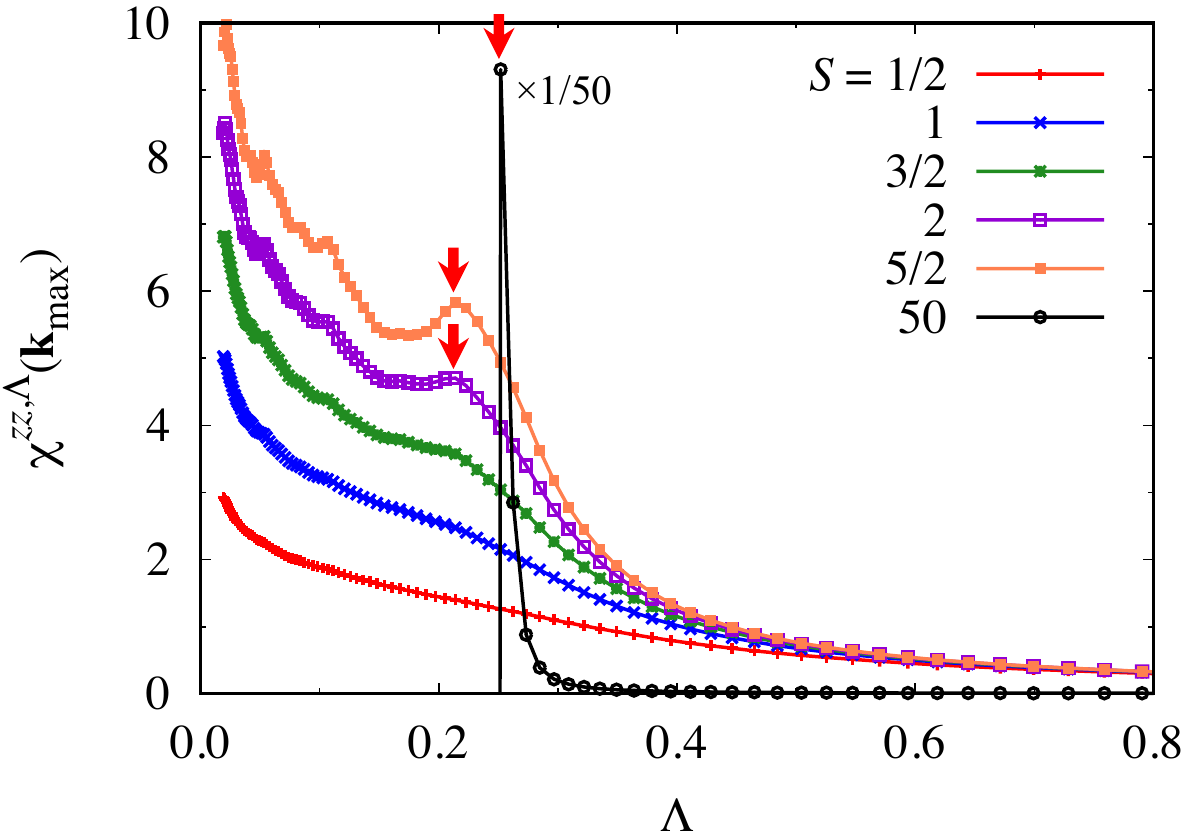}
    \caption{Spin susceptibility for the $z$-spin component, $\chi^{zz,\Lambda}(\mathbf{k}_{\mathrm{max}})$, as a function of the cutoff scale $\Lambda$ for the FM Kitaev case with $\xi=0.75$. 
$\mathbf{k}_{\mathrm{max}}$ is the wave vector at which the susceptibility becomes maximum. The red arrows indicate the characteristic cutoff scale $\Lambda^*$ at which $\chi^{zz,\Lambda}(\mathbf{k}_{\mathrm{max}})$ shows an anomaly.}
    \label{fig:flow_Kitaev}
\end{figure}

\begin{figure}
    \centering
    \includegraphics[width=1.0\columnwidth,clip]{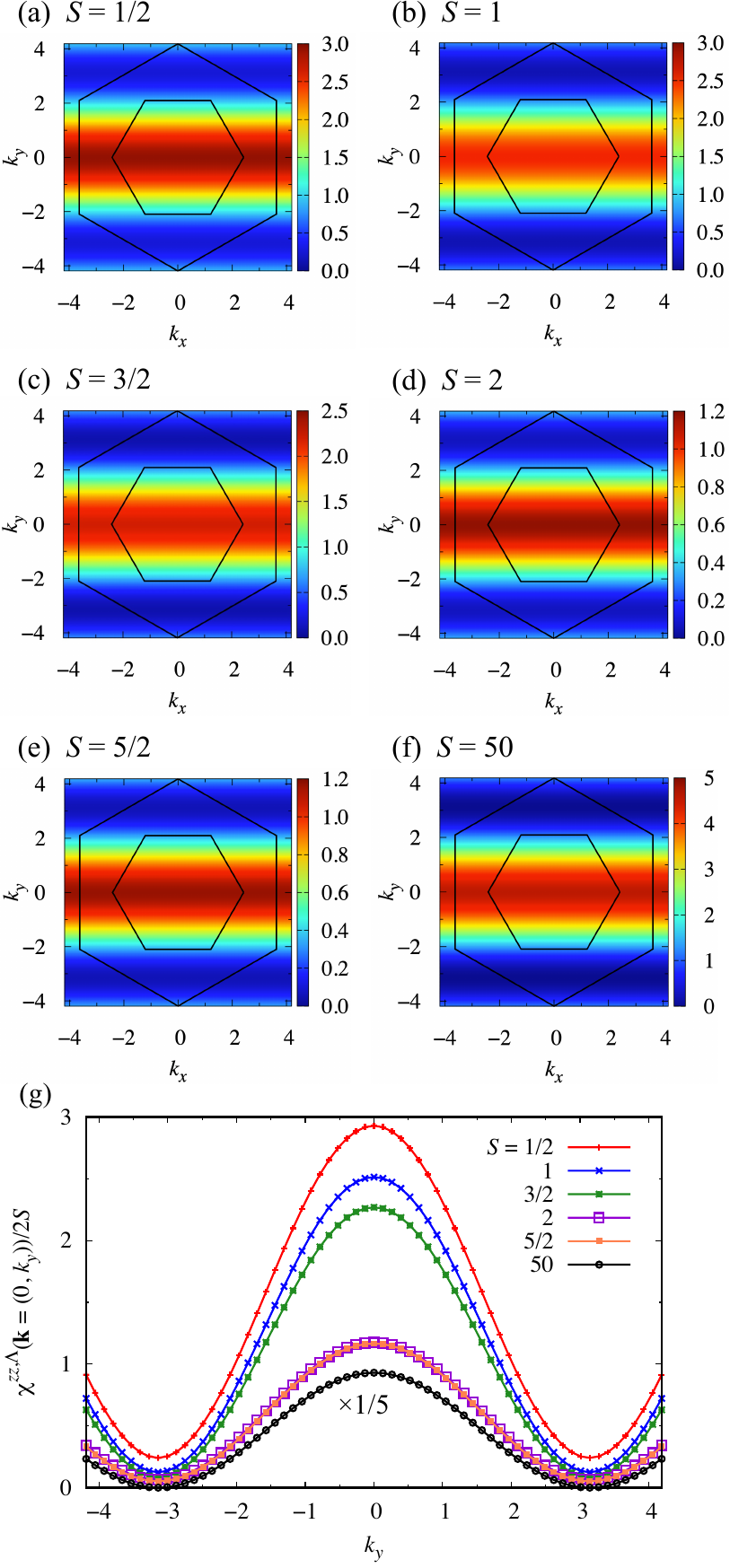}
    \caption{$\mathbf{k}$ dependences of $\chi^{zz,\Lambda}(\mathbf{k})/2S$ for the FM Kitaev case with $\xi=0.75$ for (a) $S=1/2$, (b) $S=1$, (c) $S=3/2$, (d) $S=2$, (e) $S=5/2$, and (f) $S=50$. The data in (a)--(c) are at the minimum cutoff scale $\Lambda_{\mathrm{min}}$, while those in (d)--(f) are at $\Lambda^*$ (see Fig.~\ref{fig:flow_Kitaev}). The inner black hexagon indicates the first Brillouin zone, while the outer one indicates the zone up to the third one. (g) $k_y$ dependences of $\chi^{zz,\Lambda}(\mathbf{k}=(k_x=0, k_y))/2S$. The data are plotted for the same values of $\Lambda$ in (a)--(f).
}
    \label{fig:k_map_xi0.75}
\end{figure}

Let us discuss the results for the pure Kitaev cases with $\xi=0.25$ and $0.75$. Figure~\ref{fig:flow_Kitaev} shows the $\Lambda$ dependences of $\chi^{zz, \Lambda}(\mathbf{k}_{\mathrm{max}})$ for the FM Kitaev case with $\xi=0.75$ for $S=1/2$ to $5/2$, and $50$.
Here, $\mathbf{k}_{\mathrm{max}}$ represents the wave vector at which the susceptibility becomes maximum in the reciprocal space; in the FM Kitaev case, $\mathbf{k}_{\mathrm{max}}=(k_{x}, 0)$ with arbitrary $k_{x}$, as shown in Fig.~\ref{fig:k_map_xi0.75}. 
We obtain the same results for $\chi^{zz, \Lambda}(\mathbf{k}_{\mathrm{max}})$ with $\mathbf{k}_{\mathrm{max}}=(k_{x}, \pm\pi)$ for the AFM case at $\xi=0.25$.  
The susceptibility for $S\leq 3/2$ shows no obvious anomaly down to $\Lambda_{\mathrm{min}}$, whereas it shows hump or peak like anomalies for $S\geq2$ at $\Lambda^*$ indicated by the red arrows in Fig.~\ref{fig:flow_Kitaev}.  
We plot the values of $\Lambda^{*}$ at $\xi=0.75$ and $0.25$ in Figs.~\ref{fig:Lambda_xi_diagram}(d)--\ref{fig:Lambda_xi_diagram}(f). We believe, however, that these are an artifact of the current PFFRG calculations, and that the ground state is a QSL as shown by the analytical solutions~\cite{Baskaran2008} as discussed below. 

The $\mathbf{k}$ dependences of $\chi^{zz, \Lambda}(\mathbf{k})$ for the FM Kitaev case are shown in Fig.~\ref{fig:k_map_xi0.75}. Note that $\chi^{xx, \Lambda}(\mathbf{k})$ and $\chi^{yy, \Lambda}(\mathbf{k})$ are obtained by $C_3$ rotations.  
The results indicate that $\chi^{zz, \Lambda}(\mathbf{k})$ is well approximated by $\propto \cos k_y+\mathrm{const.}$ for all $S$. This means that the spin correlations are negligible beyond nearest neighbors, and the ground state is a QSL similar to the analytical solutions~\cite{Baskaran2008}. 
Interestingly, even for $S\geq 2$ where $\chi^{zz, \Lambda}(\mathbf{k}_{\mathrm{max}})$ shows an anomaly at $\Lambda^*$,  $\chi^{zz, \Lambda}(\mathbf{k})$ has the cosine form. 
Since $\chi^{zz, \Lambda}(\mathbf{k})$ develops a sharp peak as $\Lambda \to \Lambda_{\mathrm{c}}$ for conventional magnetic orderings, the results suggest that the anomalies at $\Lambda^*$ are not due to magnetic instabilities but presumably due to numerical instabilities in the PFFRG calculations; 
this QSL-like behavior would hold for $\Lambda \to 0$ once the anomalies at $\Lambda^*$ are suppressed by better approximations, for example, by taking finer grids of $\omega$ and larger $N$, or the methods beyond the Katanin scheme.

\subsection{Around the pure Kitaev cases}\label{subsec:vicinity}

\begin{figure*}[t]
    \centering
    \includegraphics[width=1.0\linewidth,clip]{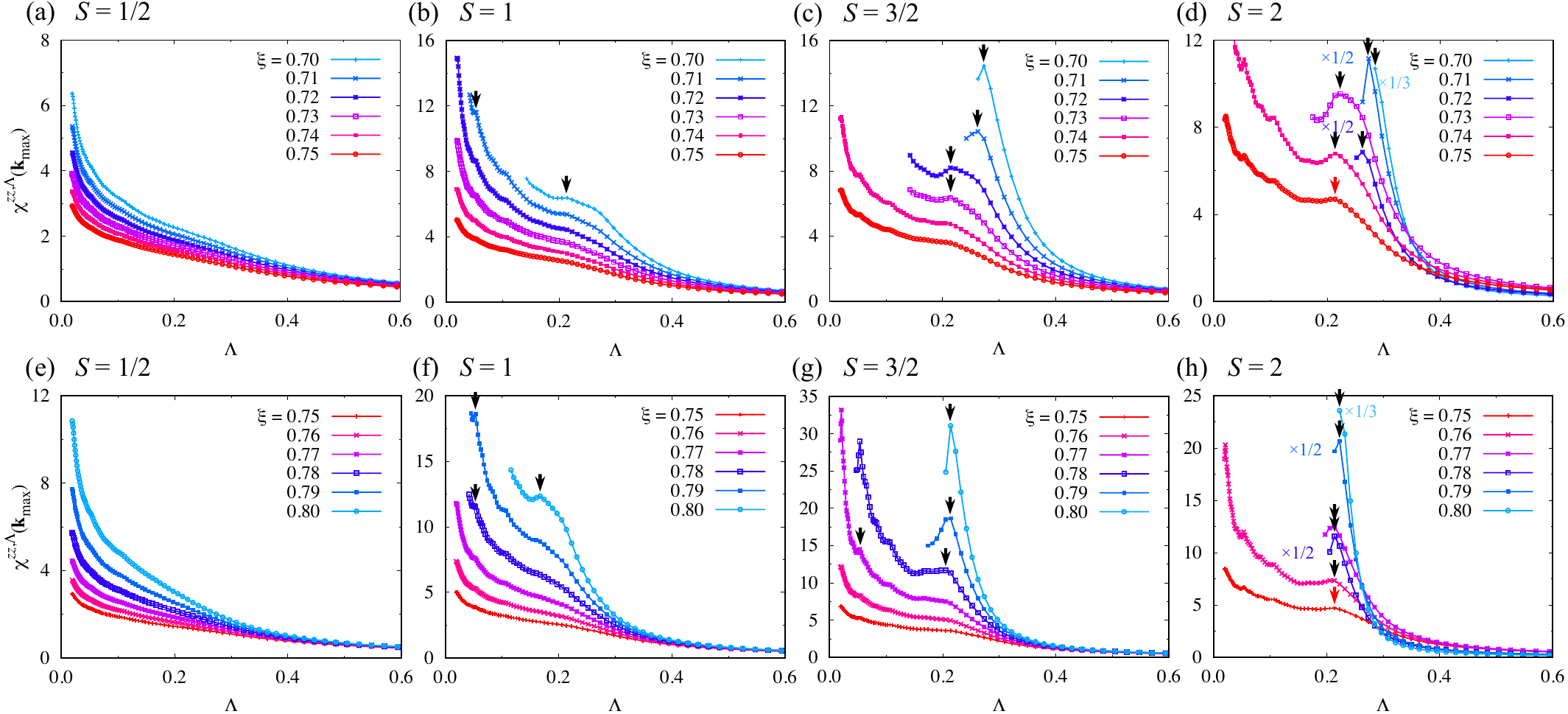}
    \caption{$\Lambda$ dependences of $\chi^{zz,\Lambda}(\mathbf{k}_{\mathrm{max}})$ in the vicinities of the FM Kitaev case for (a)(e) $S=1/2$, (b)(f) $S=1$, (c)(g) $S=3/2$, and (d)(h) $S=2$; (a)--(d) are for $0.70\leq \xi \leq 0.75$, and (e)--(h) are for $0.75\leq \xi \leq 0.80$.  
The black arrows indicate the critical cutoff scale $\Lambda_{\mathrm{c}}$, while the red ones for the data at $\xi=0.75$ in (d) and (h) denote $\Lambda^*$ as in Fig.~\ref{fig:flow_Kitaev}.
}
    \label{fig:flow_2S1to2S4_FM}
\end{figure*}

Now we discuss the effect of the Heisenberg interaction in the vicinities of the pure Kitaev cases.  
Figure~\ref{fig:flow_2S1to2S4_FM} shows the $\Lambda$ dependences of $\chi^{zz, \Lambda}(\mathbf{k}_{\mathrm{max}})$ for $S=1/2$ to $2$ around the FM Kitaev case with $\xi=0.75$. 
Here, $\mathbf{k}_{\mathrm{max}}$ is located at the $\Gamma$ point for $\xi<0.75$ and the $\mathrm{X}$ points for $\xi>0.75$, 
while $\mathbf{k}=(k_x,0)$ with arbitrary $k_x$ at $\xi=0.75$. 
We find that $\chi^{zz, \Lambda}(\mathbf{k}_{\mathrm{max}})$ does not show anomalies around $\xi=0.75$:
$0.635\leq\xi\leq0.8225$ 
for $S=1/2$, 
$0.715\leq\xi\leq0.7775$ 
for $S=1$, and 
$0.7325\leq\xi\leq0.7675$ 
for $S=3/2$, 
from which we identify the QSL phases in Fig.~\ref{fig:S_xi_diagram}. 
In contrast, $\chi^{zz, \Lambda}(\mathbf{k}_{\mathrm{max}})$ for $S=2$ shows an anomaly for all values of $\xi$ calculated near $\xi=0.75$, as shown in Figs.~\ref{fig:flow_2S1to2S4_FM}(d) and \ref{fig:flow_2S1to2S4_FM}(h). 

\begin{figure*}[t]
    \centering
    \includegraphics[width=1.0\linewidth,clip]{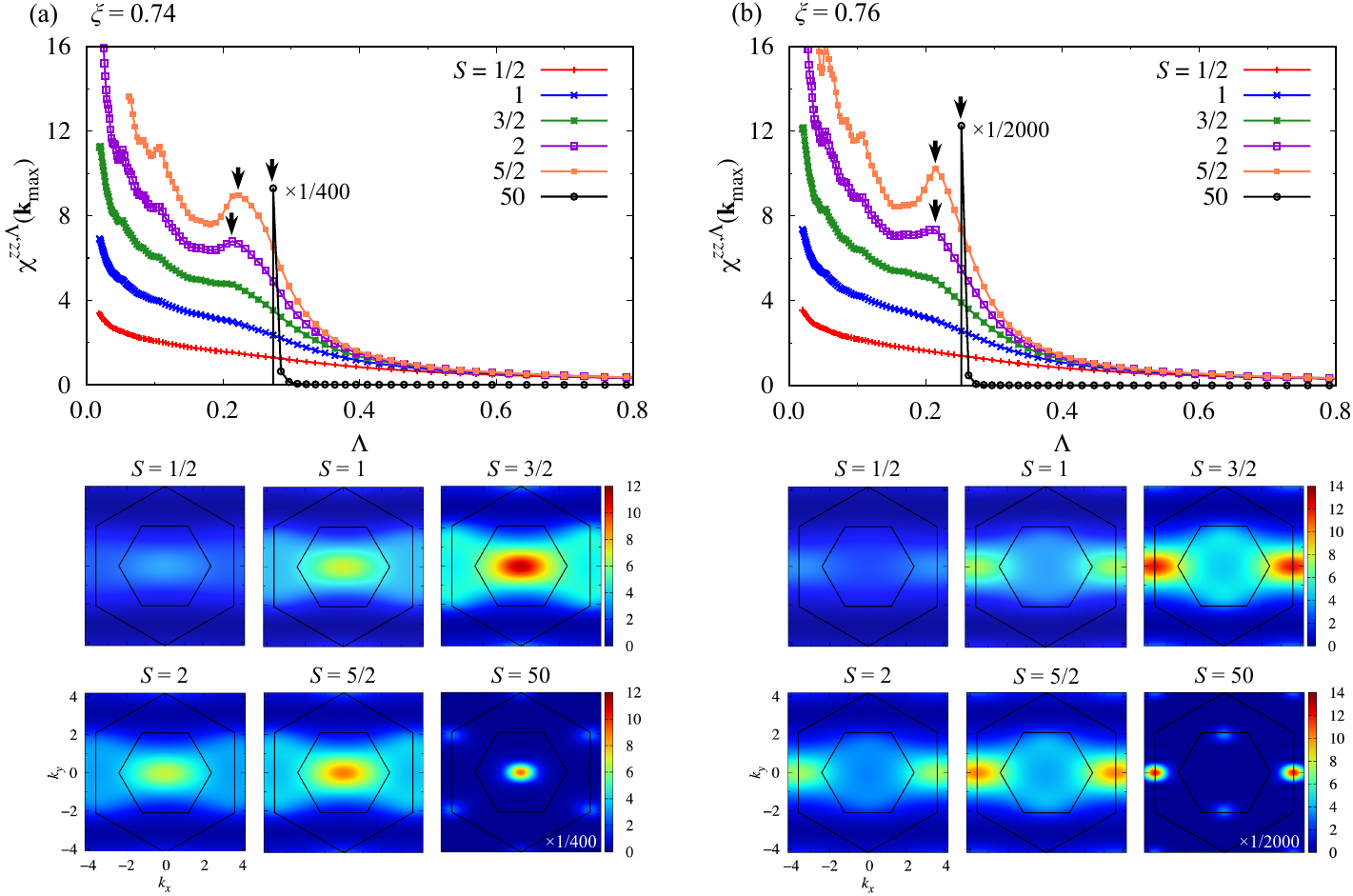}
    \caption{Behaviors of $\chi^{zz,\Lambda}(\mathbf{k})$ in the vicinity of the pure FM Kitaev case: (a) $\xi=0.74$ and (b) $\xi=0.76$. The upper panels show $\Lambda$ dependences of $\chi^{zz,\Lambda}(\mathbf{k}_{\mathrm{max}})$, and the lower panels show $\bf{k}$ dependences of $\chi^{zz,\Lambda}(\mathbf{k})$. In the lower panels, the data for $S=1/2$ to $3/2$ are at $\Lambda_{\mathrm{min}}$, while those for $S=2$, $5/2$, and $50$ are at $\Lambda_{\mathrm{c}}$ shown by the black arrows in the upper panels. 
}
    \label{fig:flow_xi0.74_0.76}
\end{figure*}

Let us closer look at the behavior of $\chi^{zz, \Lambda}(\mathbf{k})$ in the very vicinity of $\xi=0.75$. Figure~\ref{fig:flow_xi0.74_0.76} summarizes the results for (a) $\xi=0.74$ and (b) $\xi=0.76$. 
In the upper panels, $\chi^{zz, \Lambda}(\mathbf{k}_{\mathrm{max}})$ does not show clear anomalies for $S\leq3/2$, while it shows an anomaly at $\Lambda_{\mathrm{c}}$ for $S\geq2$ at both $\xi=0.74$ and $0.76$. 
The data for $S=50$ are separately shown in the logarithmic scale in Appendix~\ref{appx:log_plot}.
The lower panels of Fig.~\ref{fig:flow_xi0.74_0.76} show the $\mathbf{k}$ dependences of $\chi^{zz, \Lambda}(\mathbf{k})$. 
We present the data at $\Lambda_{\mathrm{min}}$ for $S=1/2$, $1$, and $3/2$, and those at $\Lambda_{\mathrm{c}}$ for $S=2$, $5/2$, and $50$. For all $S$, we find that $\chi^{zz, \Lambda}(\mathbf{k})$ has peaks at the wave vectors corresponding to the FM and stripy AFM states at $\xi=0.74$ and $0.76$, respectively; see Figs.~\ref{fig:KH_model}(b) and \ref{fig:KH_model}(e).  
Nevertheless, $\chi^{zz, \Lambda}(\mathbf{k})$ behaves differently for $S\leq 3/2$ and $S\geq 2$. 
In the former cases, even at $\Lambda_{\mathrm{min}}$, the peaks of $\chi^{zz, \Lambda}(\mathbf{k})$ remain broad with a remnant of the cosine spectrum in the background, reflecting the QSL ground state, while the peak intensities become large with the increase of $S$. In contrast, for the latter, $\chi^{zz, \Lambda}(\mathbf{k})$ at $\Lambda=\Lambda_{\mathrm{c}}$ already develop rather strong peaks, which we regard as the instabilities to magnetic orderings.
From these results, we conclude that the QSL phases are stable around $\xi=0.75$ in the cases of $S=1/2$, $1$, and $3/2$, while the regions are reduced rapidly  
by the increase of $S$ and become vanishingly small for $S=2$.

\begin{figure*}[t]
    \centering
    \includegraphics[width=1.0\linewidth,clip]{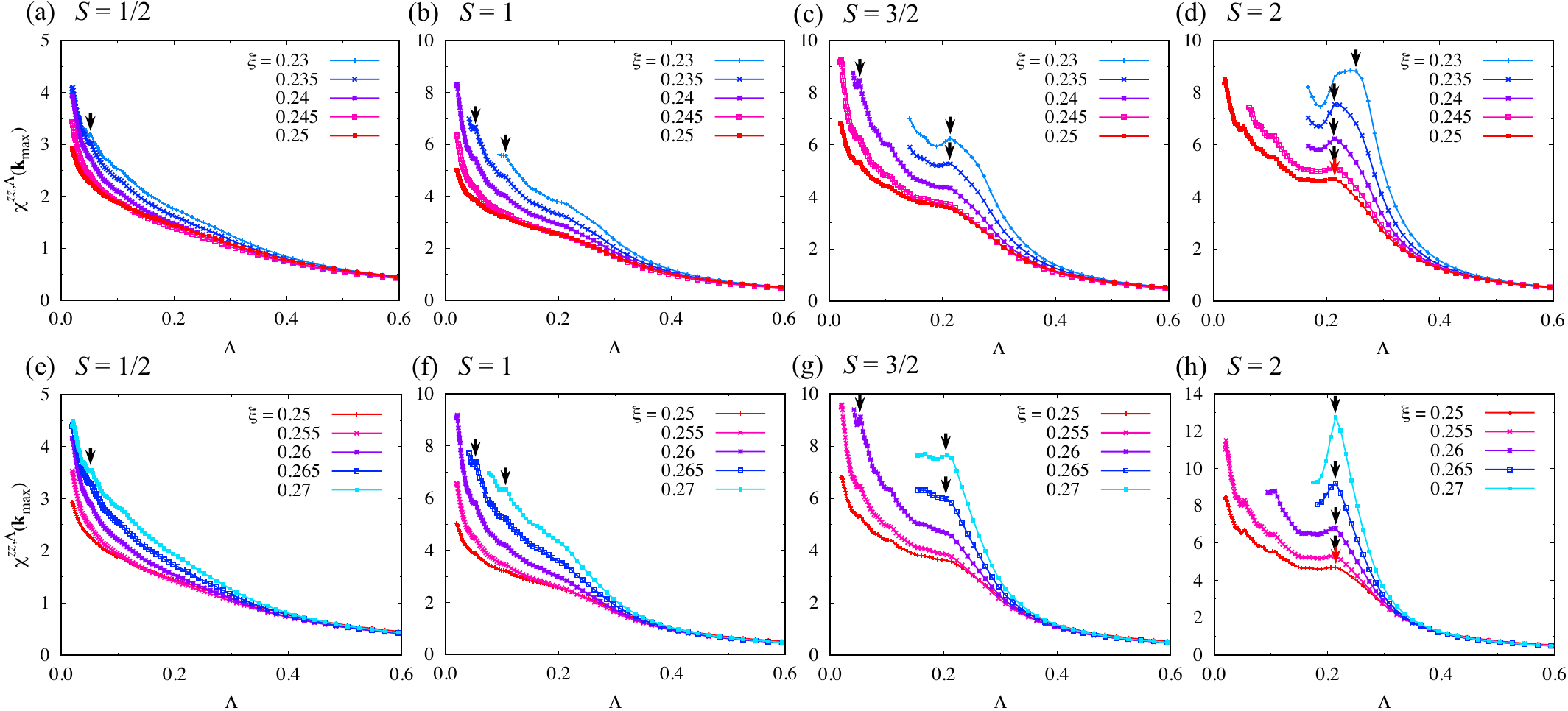}	
    \caption{Similar plots to Fig.~\ref{fig:flow_2S1to2S4_FM} in the vicinities of the AFM Kitaev case: (a)--(d) $0.23\leq \xi\leq 0.25$ and (e)--(h) $0.25\leq \xi\leq0.27$. 
}
    \label{fig:flow_2S1to2S4_AFM}
\end{figure*}

Similarly, around the AFM Kitaev case with $\xi=0.25$, we identify the QSL phases for $S\leq 3/2$, which become vanishingly small for $S\geq 2$. 
The results are shown in Fig.~\ref{fig:flow_2S1to2S4_AFM}. 
In this case, $\mathbf{k}_{\mathrm{max}}$ is located at the $\Gamma'$ points for $\xi<0.25$, and the $\mathrm{Y}$ points for $\xi>0.25$, while $\mathbf{k}=(k_x,\pm\pi)$ with arbitrary $k_x$ at $\xi=0.25$.
 $\Lambda$ dependence of $\chi^{zz, \Lambda}(\mathbf{k}_{\mathrm{max}})$ does not show anomalies around $\xi=0.25$:
$0.2325\leq\xi\leq0.2675$ for $S=1/2$, $0.2375\leq\xi\leq0.2625$ for $S=1$, and $0.2425\leq\xi\leq0.2575$ for $S=3/2$, from which we identify the QSL phases in Fig.~\ref{fig:S_xi_diagram}.
In contrast, it shows clear anomaly in the case of $S=2$ for 
all values of $\xi$ calculated around $\xi=0.25$, as shown in Figs.~\ref{fig:flow_2S1to2S4_AFM}(d) and \ref{fig:flow_2S1to2S4_AFM}(h). 

In our results, the QSL phases become vanishingly small for $S\geq2$ in both FM and AFM cases, even though their widths in the phase diagram for $S\leq3/2$ are different.
Considering the overestimates for the $S=1/2$ and $1$ cases in Fig.~\ref{fig:S_xi_diagram} compared with the previous results by other methods, we believe that the strong suppression of the QSL for $S\geq2$ is valid, while more precise estimate of the phase boundaries needs further efforts, especially for large $S$.

\subsection{Magnetically ordered states}\label{subsec:ordered}

\begin{figure}
    \centering
    \includegraphics[width=0.9\columnwidth,clip]{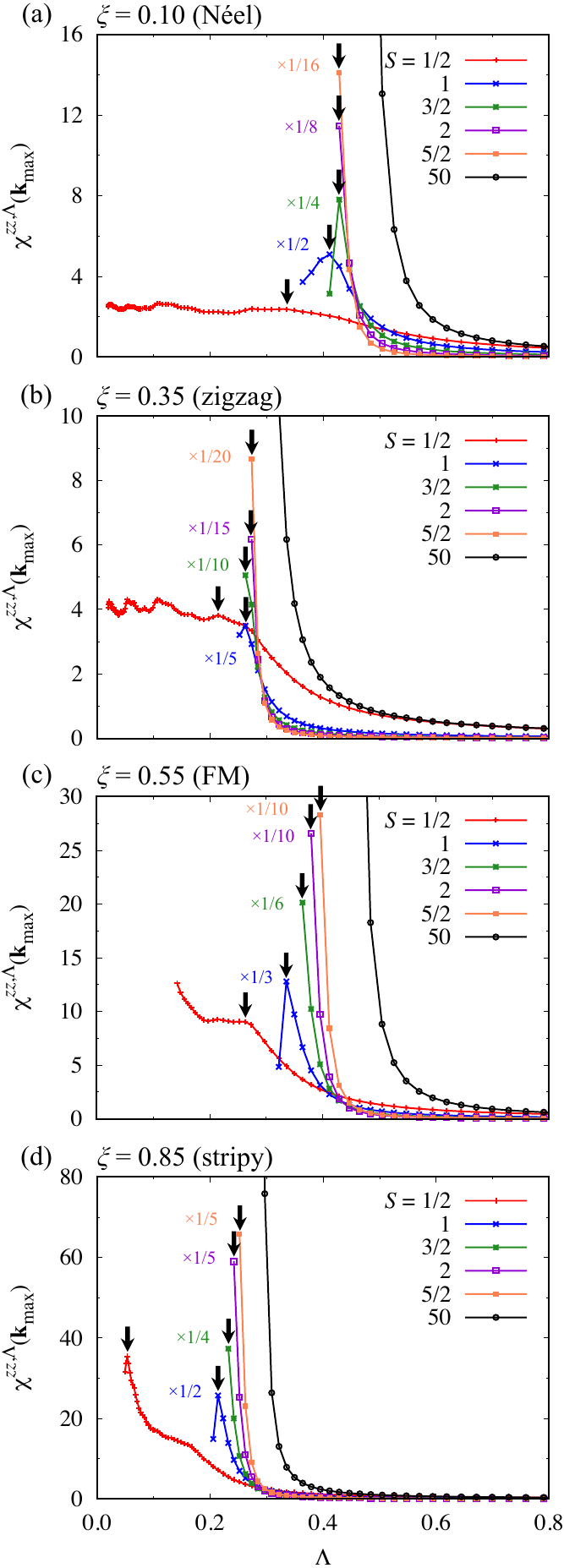}
    \caption{$\Lambda$ dependences of $\chi^{zz,\Lambda}(\mathbf{k}_{\mathrm{max}})$ at (a) $\xi=0.10$, (b) $\xi=0.35$, (c) $\xi=0.55$, and (d) $\xi=0.85$.  
The black arrows indicate $\Lambda_{\mathrm{c}}$. 
} 
    \label{fig:flow_ordered}
\end{figure}

\begin{figure}
    \centering
    \includegraphics[width=1.0\columnwidth,clip]{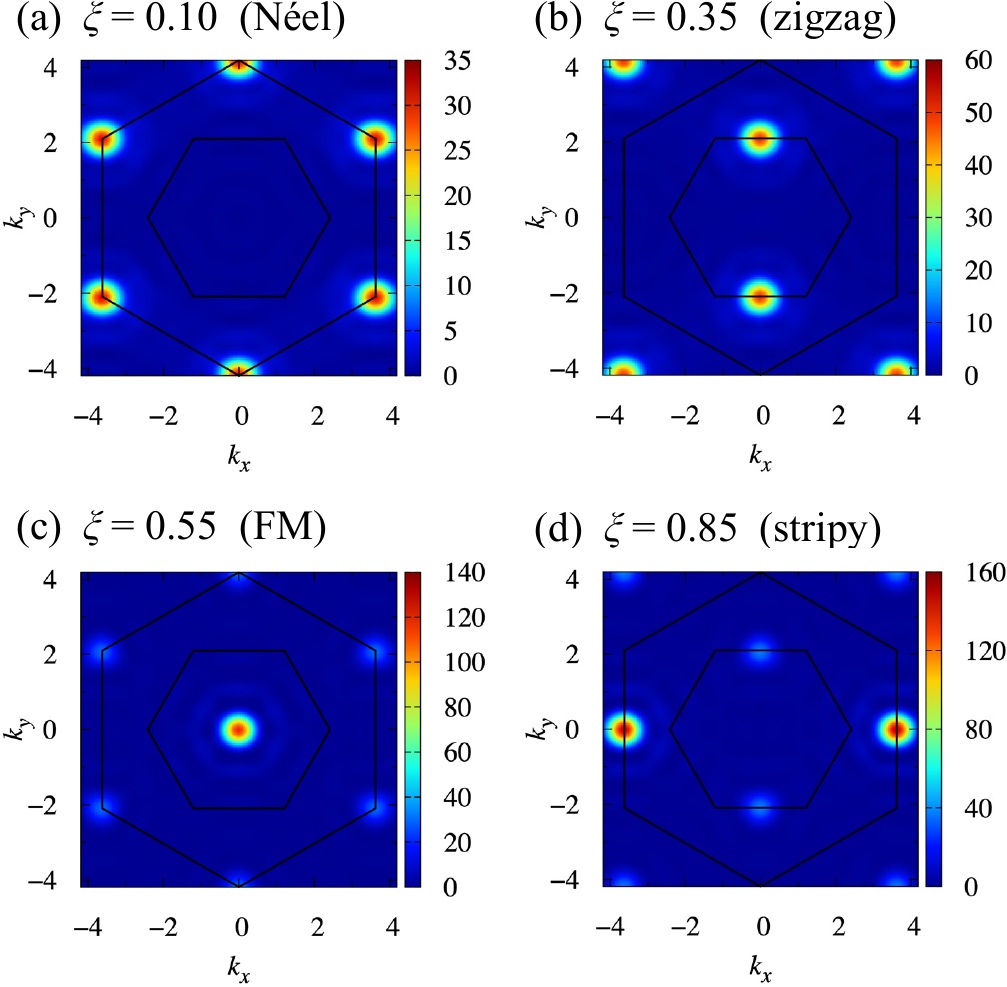}
    \caption{$\mathbf{k}$ dependences of $\chi^{zz,\Lambda}(\mathbf{k})$ for the case with $S=3/2$ at (a) $\xi=0.10$, (b) $\xi=0.35$, (c) $\xi=0.55$, and (d) $\xi=0.85$. The data are at the critical cutoff scale $\Lambda_{\mathrm{c}}$ indicated by the arrows in Fig.~\ref{fig:flow_ordered}.
}
    \label{fig:k_map_ordered}
\end{figure}

Finally, we present the results for the magnetically ordered states.  
Figure~\ref{fig:flow_ordered} shows the $\Lambda$ dependences of $\chi^{zz, \Lambda}(\mathbf{k}_{\mathrm{max}})$ for four ordered phases: (a) N\'eel AFM ($\xi=0.10$), (b) zigzag AFM ($\xi=0.35$), (c) FM ($\xi=0.55$), and (d) stripy AFM ($\xi=0.85$). 
See the spin patterns in Figs.~\ref{fig:KH_model}(b)--\ref{fig:KH_model}(e).
$\mathbf{k}_{\mathrm{max}}$ is located at the $\Gamma'$ points for $\xi=0.10$, the $\mathrm{Y}$ points for $\xi=0.35$, the $\Gamma$ point for $\xi=0.55$, and the $\mathrm{X}$ points for $\xi=0.85$.
We find that $\chi^{zz, \Lambda}(\mathbf{k}_{\mathrm{max}})$ shows anomalies at $\Lambda_{\mathrm{c}}$ indicated by the black arrows for all $S$. We plot the data for $S=50$ separately in the logarithmic scale in Appendix~\ref{appx:log_plot}. 
These anomalies indicate the magnetic instabilities in the four regions. We note that the values of $\Lambda_{\mathrm{c}}$ increase with $S$, as discussed in Fig.~\ref{fig:Lambda_xi_diagram}.

Figure~\ref{fig:k_map_ordered} presents the typical data of $\chi^{zz, \Lambda_{\mathrm{c}}}(\mathbf{k})$ at the same values of $\xi$ as in Fig.~\ref{fig:flow_ordered}, by taking examples of $S=3/2$. We find distinct peaks that develop at $\mathbf{k}_{\mathrm{max}}$, corresponding to each magnetic ordering. We obtain qualitatively similar results for the other values of $S$. 
The results are consistent with the previous ones for $S=1/2$, $1$, and $50$~\cite{Chaloupka2013, Dong2020, Price2013}.

\section{Summary}\label{sec:summary}
To summarize, we have studied the spin-$S$ Kitaev-Heisenberg model, which we consider to be one of the minimal models for the higher-spin candidates for the Kitaev magnets, by using the PFFRG method extended to general $S$. We elucidated the ground-state phase diagram systematically by changing $S$ and the ratio between the Kitaev and Heisenberg interactions.  
We obtained QSL behaviors in the pure Kitaev cases without the Heisenberg interaction in both FM and AFM cases for all $S$, consistent with the analytical solutions~\cite{Baskaran2008}. 
We found that, beyond the previous studies for $S=1/2$ and $1$, the QSL phases around the pure Kitaev cases are rapidly reduced by increasing $S$ and the regions become vanishingly small for $S\geq 2$, while the other magnetically ordered phases remain robust.

Our results indicate that quantum fluctuations are essential to preserve the Kitaev QSL against the Heisenberg interaction, and it would be hard to find a good candidate material for $S\geq 2$ without very fine tuning of the interaction parameters. 
For $S=3/2$, a candidate CrI$_{3}$ was claimed to be close to the Kitaev QSL: $\xi$ in Eq.~\eqref{eq:KandJ} was estimated as $\simeq 0.762$ 
by the angle-dependent ferromagnetic resonance experiment~\cite{Lee2020a},  
which lies, in our calculations, in the QSL phase around the FM Kitaev cases as discussed in Sec.~\ref{subsec:vicinity}. In reality, however, CrI$_{3}$ exhibits FM ordering at low temperature, which was ascribed to the effect of other interactions like a symmetric off-diagonal interaction called the $\Gamma$ term~\cite{Lee2020a}. 
This suggests that the Kitaev QSL regions are further reduced by including other interactions. Such investigation by systematically changing $S$ is left for future study.

\begin{acknowledgments}
K.F. thanks Yusuke Kato for constructive suggestions.
The authors thank T. Misawa and T. Okubo for fruitful discussions.
Parts of the numerical calculations have been done using the facilities of the Supercomputer Center, the Institute for Solid State Physics, the University of Tokyo, the Information Technology Center, the University of Tokyo, and the Center for Computational Science, University of Tsukuba.
This work was supported by Japan Society for the Promotion of Science (JSPS) KAKENHI Grant Nos. 19H05825 and 20H00122.
K.F. was supported by the Program for Leading Graduate Schools (MERIT).
\end{acknowledgments}
\appendix
\section{Effect of frequency discretization}\label{appx:stepwise}

\begin{figure}
    \centering
    \includegraphics[width=1.0\columnwidth,clip]{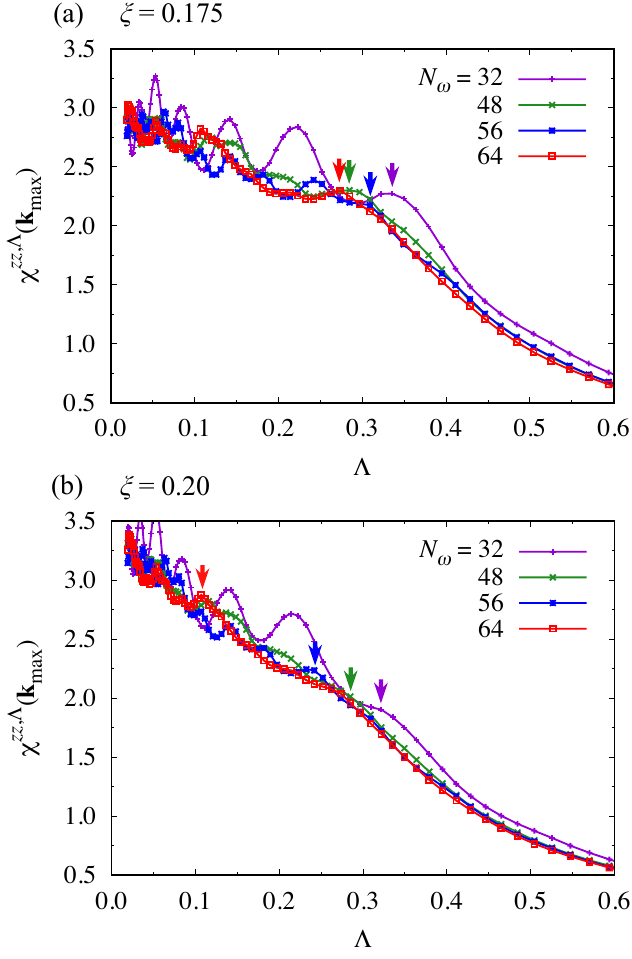}
    \caption{$\Lambda$ dependence of $\chi^{zz,\Lambda}(\mathbf{k}_{\mathrm{max}})$ for the $S=1/2$ case at (a) $\xi=0.175$ and (b) $\xi=0.20$. The data are plotted for different numbers of the frequency grid from $N_{\omega}=32$ to $64$. 
The colored arrows indicate $\Lambda_{\mathrm{c}}$ for each data. 
}
    \label{fig:flow_Nw_xi0.175_0.20}
\end{figure}

In this Appendix, we discuss the effect of the discretization of $\omega$ in the PFFRG calculations. 
Figure~\ref{fig:flow_Nw_xi0.175_0.20} shows $\chi^{zz, \Lambda}(\mathbf{k}_{\mathrm{max}})$ for different numbers of $\omega$ grids, $N_\omega$, in the case of the $S=1/2$ Kitaev-Heisenberg model at $\xi=0.175$ and $0.20$ in the N\'eel AFM region. 
Here, we discretize the frequency range of $10^{-4}\leq\omega\leq250$ logarithmically with $N_{\omega}$ frequency points. The system size and the $\Lambda$ grids are the same as in the main text. 
We find that the value of $\Lambda_{\mathrm{c}}$ at which $\chi^{zz, \Lambda}(\mathbf{k}_{\mathrm{max}})$ shows an anomaly varies with $N_{\omega}$; in particular, it varies non-monotonically for $\xi=0.175$. 
In addition, between $\xi=0.175$ and $\xi=0.20$, $\Lambda_{\mathrm{c}}$ for $N_{\omega}=56$ and $64$ take different values, whereas those for $N_{\omega}=32$ and $48$ are the same. These results show that the estimate of $\Lambda_{\mathrm{c}}$ is sensitive to $N_\omega$. 
We thus speculate that the stepwise behavior of $\Lambda_{\mathrm{c}}$ in Fig.~\ref{fig:Lambda_xi_diagram} is due to the discretization of $\omega$, and expect that $\Lambda_{\mathrm c}$ behaves more smoothly for larger $N_\omega$. 
Meanwhile, the reason why $\Lambda_{\mathrm{c}}$ varies smoothly for large $S$ 
is that the frequency dependence becomes more irrelevant for less quantum fluctuations~\cite{Baez2017}. 

\section{Log plot of the susceptibility}\label{appx:log_plot}

\begin{figure}
    \centering
    \includegraphics[width=1.0\columnwidth,clip]{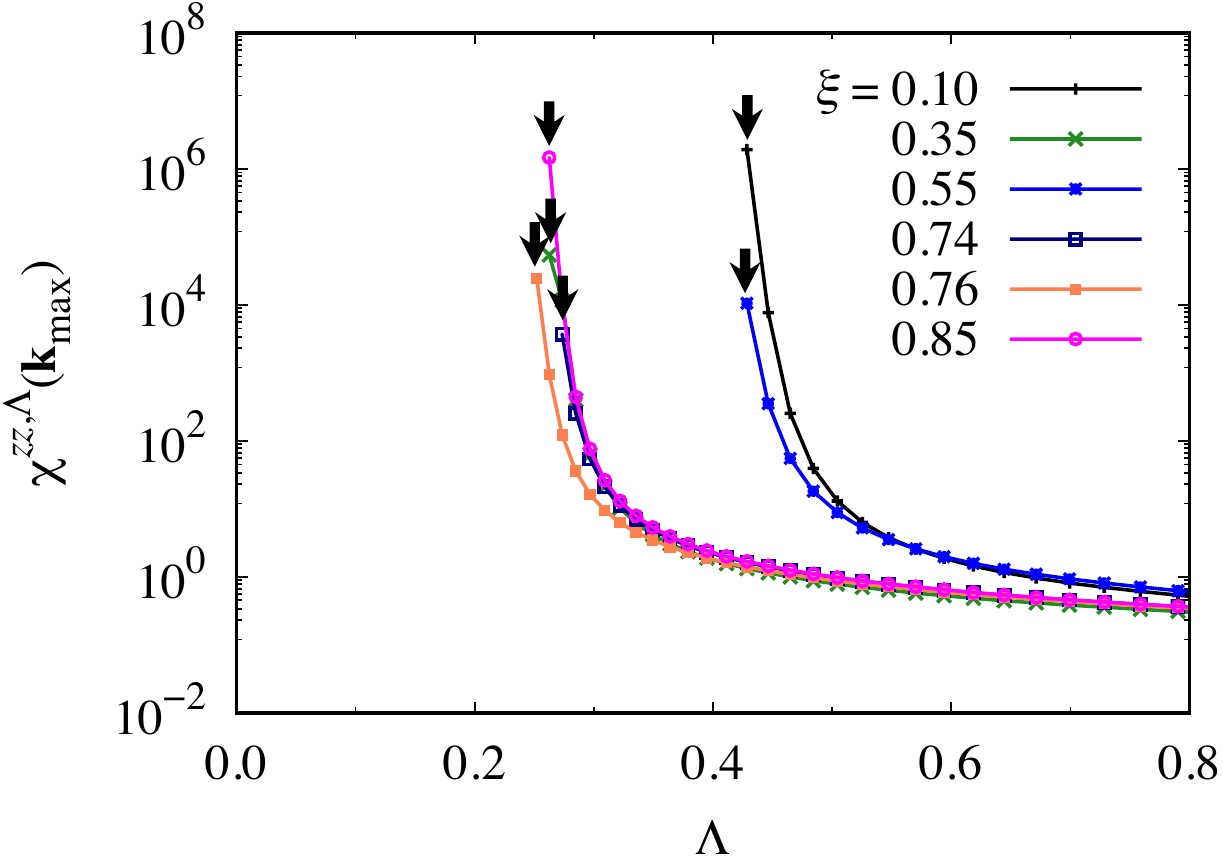}
    \caption{Log plot of $\chi^{zz,\Lambda}(\mathbf{k}_{\mathrm{max}})$ as a function of $\Lambda$ for $S=50$ 
    at $\xi=0.10$, 0.35, 0.55, 0.74, 0.76, and 0.85. The black arrows indicate $\Lambda_{\mathrm{c}}$.
    }
    \label{fig:flow_log_S50}
\end{figure}

Figure~\ref{fig:flow_log_S50} plots $\chi^{zz,\Lambda}(\mathbf{k}_{\mathrm{max}})$ for $S=50$ in Figs.~\ref{fig:flow_xi0.74_0.76} and \ref{fig:flow_ordered} in the logarithmic scale. 
All the data show very rapid increases as $\Lambda \to \Lambda_{\mathrm{c}}$.


\bibliography{library}

\end{document}